\documentclass{edp-conf}
\usepackage{graphicx}
\newcommand{\dfr}{d\raise0.3ex\hbox{\kern-0.5ex\char"013 }} %fract deriv.
\begin{document}

\TitreGlobal{SF2A 2004}

%%-----------------------------
%%      the top matter
%%-----------------------------
\title{Generalized macroscopic Schr\"odinger equation in scale relativity}
\author{C\'el\'erier, M.N.}\address{Laboratoire Univers et TH\'eories (LUTH), Observatoire de Paris-Meudon, 5 place Jules Janssen 92195 Meudon Cedex, France}
\author{Nottale, L.$^1$}
\runningtitle{Schr\"odinger equation in scale relativity }
\setcounter{page}{237}
% Keep this line, even if the page will be settled afterwards..
\index{C\'el\'erier, M.N.}
\index{Nottale, L.}
% Repeat the authors here, this will help to make the final index

\maketitle
\begin{abstract}
The scale transformation laws produce, on the motion equations of 
gravitating bodies and under some peculiar assumptions, effects which are 
anologous to those of a "macroscopic quantum mechanics". When we consider 
time and space scales such that the description of the trajectories of these 
bodies (planetesimals in the case of planetary system formation, interstellar 
gas and dust in the case of star formation, etc...) is in the shape of 
non-differentiable curves, we obtain fractal curves of fractal dimension 2. 
Continuity and non-differentiability yield a fractal space and a symmetry 
breaking of the differential time element which gives a doubling of the 
velocity fields. The application of a geodesics principle leads to motion 
equations of Schr\"odinger-type. When we add an outside gravitational field, we 
obtain a Schr\"odinger-Poisson system. We give here the derivation of the 
Schr\"odinger equation for chaotic systems, i.e., with time scales much longer 
than their Lyapounov chaos-time. 
\end{abstract}
%
%%-----------------------------
%%      your text
%%-----------------------------
\section{Introduction: the foundations of scale relativity}
 
The scale relativity theory is a geometric representation of nature (as 
General Relativity is a geometric representation of gravitation) based on 
a continuity hypothesis and constrained by the relativity principle. It 
includes in its description non-differentiable manifolds, thus the fractal character 
of space-time in the general meaning: fractal $\equiv {\cal L}_{D_T}
(\epsilon){\longrightarrow \atop{\epsilon\rightarrow 0}}\infty$ (for the 
demonstration see (Nottale, 1993)). 

This approach involves a scale dependence of the reference frames. We therefore 
add to the standard variables (position, orientation and motion), 
which characterize the reference frame, other variables characterizing its 
scale state. The use of differential equations is made possible thanks to 
the representation of physical quantities, usually mere functions 
of the space-time coordinates $f(x)$, by fractal functions $f[x(\epsilon), 
\epsilon]$, explicitly depending on the scale variables, generically noted 
$\epsilon$. 

Generalizing the definition of fractal functions to fractal space(-time)s 
(fractal space(-time) $\equiv$ equivalence class of a family of Riemannian 
space(-time)s), we obtain scale dependent geodesics equations  
and, therefore, an infinite family of geodesics.

\section{Dynamics in scale relativity}

In the scale relativity theory, the Schr\"odinger equation is derived from 
three fundamental conditions which are consequences of the non-differentiability:

(1) The fractality of space, which implies that the number of 
geodesics is infinite. We are therefore led 
to use a fluid-like description where the velocity $v(t)$ is replaced, as a first step, by 
a velocity field $v[x(t),t]$. 

(2) The fractal geometry of each geodesic, which implies that the velocity field is actually a
fractal function, $V[x(t,dt),t,dt]$, explicitly depending on a scale 
variable, identified, in the present case, to the differential element 
$dt$. One can show (Nottale, 1993) that it can be decomposed in terms of the sum of a classical (differentiable) velocity field and of a divergent fluctuation field,
\begin{equation}
V[x(t,dt),t, dt] = v[x(t),t] + w[x(t,dt),t,dt] = v \left[1 + \zeta 
\left({\tau\over dt}\right)^{1-1/D_{F}} \right],
\label{22}
\end{equation}
where $D_F$ is the fractal dimension of the geodesics. The $w$ function is a fractal fluctuation which is described in terms of a stochastic variable such that (for the critical case $D_F = 2$)
 \begin{equation}
 <w_i> = 0 \qquad\qquad <w_i w_j> = \delta_{ij}\left(2{\cal D}\over{dt}\right).
\end{equation}

(3) The non-differentiability of space, which breaks the local reflection 
invariance of the time differential element $dt$. As a result, two fractal velocity fields $V_+$ and $V_-$are defined, which are fractal functions of the scale 
variable $dt$. Each velocity field split, as in Eq.(\ref{22}), into
\begin{equation}
V_+ = v_+[x(t),t] + w_+[x(t,dt),t,dt], \quad
V_- = v_-[x(t),t] + w_-[x(t,dt),t,dt].
\end{equation}

\subsection{Covariant derivative operator}

Even after the transition to the ``classical'' 
domain is completed, there is no reason for the two velocities $v_+$ and $v_-$ 
to be equal. The natural choice for a mathematical representation of this twin-process is the use of complex numbers (C\'el\'erier and Nottale, 2004). The elementary 
displacement for each of the two processes, $dX_{\pm}$, can thus be written as 
the sum of a scale independent ``classical'' term and a fluctuation around this 
term,
\begin{equation}
dX_+(t) = v_+\;dt + d\xi_+(t), \quad dX_-(t) = v_-\;dt + d\xi_-(t).
\end{equation}
Two ``classical'' derivative $d/dt_+$ and $d/dt_-$are defined , which are applied
to the position vector $x$ to obtain the two ``classical'' velocities,
\begin{equation}
\frac{d}{dt_+}x(t)=v_+  \quad \frac{d}{dt_-}x(t) = v_-.
\end{equation}
To recover local reversibility of the time differential element, the two 
derivatives are combined in terms of a complex derivative operator,
\begin{equation}
\frac{\dfr}{dt} = {1\over 2} \left( \frac{d}{dt_+} + \frac{d}{dt_-} \right) 
- {i\over 2} \left(\frac{d}{dt_+} - \frac{d}{dt_-}\right),
\end{equation}
which, when it is applied to the position vector, gives a complex velocity,
\begin{equation}
{\cal V} = \frac{\dfr}{dt} x(t) = V -i U = \frac{v_+ + v_-}
{2} - i \;\frac{v_+ - v_-}{2}.
\end{equation}
Now, the total derivative with respect to $t$ of a function 
$f(x,t)$ contains finite terms up to the highest order. For 
a fractal dimension $D_F = 2$, it writes
\begin{equation}
{df\over {dt}} = \frac{\partial f}{\partial t} + \nabla f . {dX\over {dt}} + 
\frac{1}{2} \frac{\partial ^2 f}{\partial x_i \partial x_j} {dX_i dX_j \over {dt}}.
\end{equation}
The ``classical'' scale independent part of the term $dX_i dX_j /dt$ is finite and 
equal to $<d \xi_i\; d\xi_j> /dt = \pm 2 \; {\cal D} \; \delta _{ij}$. The last term of the scale independent part of this equation is therefore a Laplacian, and the final expression for the complex time derivative operator is derived (Nottale, 1993)
\begin{equation}
\frac{\dfr}{dt} = \frac{\partial}{\partial t}Ñ + {\cal V}. 
\nabla - i {\cal D} \Delta \;\;.
\end{equation}

%%%%%%%%%%%%%%%%%
\subsection{Improving the covariant tool of scale relativity}
\label{s:covariant}

This operator is a linear combination of first order and second order derivatives, so that its Leibniz rule is also a linear combination of the first order and second order Leibniz rules. Now the covariant character of this tool can be improved by introducing a  `symmetric product' (Pissondes, 1999) in terms of which the first order form is recovered. Another solution consists of defining a complex velocity operator, whose non-relativistic version is (Nottale, 2004)
\begin{equation}
\widehat{\cal V}= {\cal V}-i \,{\cal D}\;  \nabla \; .
\end{equation}
The covariant derivative is now written as an expression that keeps the standard (first order) form of the decomposition of a total derivative into partial derivatives, namely
\begin{equation}
\frac{\dfr}{dt}= \frac{\partial}{\partial t} +\widehat{\cal V} . \nabla\; .
\end{equation}
More generally, one defines the operator:
\begin{equation}
\widehat{\frac{\dfr f}{dt}}=\frac{\dfr f}{dt} -i {\cal D} \; \nabla f . \nabla\;.
\end{equation}
The covariant derivative of a product now writes
${\dfr (f g)}/{dt}=g\;\widehat{{\dfr f}/{dt}} +f\; \widehat{{\dfr g}/{dt}}$
i.e., one recovers the form of the first order Leibniz rule for products. Thanks to this formal tool, the standard form of the equations is preserved, i.e., a full  covariance under the generalized transformations considered here is ensured.

\subsection{Newton-Schr\"odinger equation}

Standard classical mechanics can now be generalized using this covariant tool.  The application of a Lagrangian 
formalism yields the scale relativistic Euler-Lagrange equations (Nottale, 1993), 
i.e., 

(1) For the case of inertial motion, a geodesics equation: $\dfr {\cal V} /dt = 0$. 

(2) For the case when the external structuring field is a scalar potential $\Phi$, a 
Newton-type equation of dynamics: $m \dfr {\cal V} /dt = - \nabla \Phi$. 

A complex wave function is introduced, which is another expression for the 
complex action ${\cal S}$: $\psi = e^{i{\cal S}/{\cal S}_{0}}$. We substitute it 
into the Euler-Lagrange equation, as well as the complex velocity which 
is the gradient of the complex action: $ {\cal V} = \nabla {\cal S}/ m$. The choice ${\cal S}_{0}=2 m {\cal D}$ finally allows to write this 
equation as a gradient, which, after integration, yields the Newton-Schr\"odinger equation (Nottale, 1993),
\begin{equation}
{\cal D}^2 \Delta \psi + i {\cal D} \frac{\partial}{\partial t} \psi = \frac{\Phi}{2m}\psi \;.
\end{equation}

\section{Conclusion}

We have recalled how, under three general conditions involving 
non-differentiability and fractality, the fundamental equation of dynamics can be 
transformed to take the form of a generalized Schr\"odinger equation. Such an equation is naturally structuring, since, once the potential and the symmetry and limiting conditions are specified, its solutions yield probability densities that describe the tendency for the system to make structures (Nottale, 1997). Various applications of this approach are given in other contributions to the present issue. 

%%-----------------------------
%%      your bibliography
%%-----------------------------

\end{document}